# Magnetic ordering in the $J_{\text{eff}}$ = 0 Nickelate NiRh$_2$O$_4$ prepared via a solid-state metathesis

Yuya Haraguchi[1,†], Daisuke Nishio-Hamane[2] and Hiroko Aruga Katori[1]
[1]Department of Applied Physics and Chemical Engineering, Tokyo University of Agriculture and Technology, Koganei, Tokyo 184-8588, Japan
[2]The Institute for Solid State Physics, The University of Tokyo, Kashiwa, Chiba 277-8581, Japan
†chiyuya3@go.tuat.ac.jp

In spinel-type nickelate NiRh$_2$O$_4$, magnetic ordering is observed upon the sample synthesized via kinetically controlled low-temperature solid-state metathesis, as opposed to previously-reported samples obtained through conventional solid-state reaction. Our findings are based on a combination of bulk susceptibility and specific heat measurements that disclose a Néel transition temperature of $T_{\text{N}}$ = 45 K in this material, which might feature spin-orbit entanglement in the tetragonally-coordinated $d^8$ Mott insulators. The emergence of magnetic ordering upon alteration of the synthesis route indicates that the suppression of magnetic ordering in the previous sample was rooted in the cation-mixing assisted by the entropy gain that results from high-temperature reactions. Furthermore, the $J_{\text{eff}}$ = 0 physics, instead of solely the spin-only $S$ = 1, describes the observed enhancement of effective magnetic moment well. Overseeing all observations and speculations, we propose that the possible mechanism responsible for the emergent magnetic orderings in NiRh$_2$O$_4$ is the condensation of $J_{\text{eff}}$ = 1 exciton, driven by the interplay of the tetragonal crystal field and superexchange interactions.

## I. Introduction

The phenomenon driven by spin-orbit coupling (SOC) is a fundamental aspect of various disciplines within quantum physics, including but not limited to quantum spin liquids [1-5], topological physics [6-8], multipolar order [9-12], and so on [13]. The revelation of the decisive role of SOC in stabilizing the Mott state of the layered perovskite Sr$_2$IrO$_4$ has sparked significant interest in 4$d$ and 5$d$ compounds [14,15]. The intrinsic effective orbital moment $L_{\text{eff}}$ = 1 of Ir$^{4+}$ ions, which possess $d^5$ electrons in the $t_{2g}$ manifold, renders the $J_{\text{eff}}$ = 1/2 pseudospin state a good quantum number [14,15]. Subsequent theoretical investigations have led to the seminal proposal that the Kitaev model, a quantum many-body model that realizes a quantum spin liquid as an exact solution, can be realized by harnessing the direction-dependent Ising interactions generated between $J_{\text{eff}}$ = 1/2 pseudospins on a honeycomb lattice [16]. As a result, $J_{\text{eff}}$ = 1/2 honeycomb magnets have been extensively investigated to realize spin fractionalization within the Kitaev spin liquid [17,18].

Furthermore, there is a wide range of spin-orbit-entangled states of interest among 4$d$ and 5$d$ transition metal compounds beyond $J_{\text{eff}}$ = 1/2 physics [13]. The impact of SOC on magnetism is particularly pronounced in compounds of 4$d$ and 5$d$ ions with a $d^4$ configuration due to SOC binding $S$ and $L$ moments into a local singlet state with zero total angular momentum $J_{\text{eff}}$ = 0. These nominally "nonmagnetic" ions may develop collective magnetism through excited states. Despite the absence of pre-existing local moments in the ionic ground state, $J_{\text{eff}}$ = 1 excitations become dispersive modes in a crystal, and these mobile spin-orbit excitons may condense into a magnetically ordered state [19-22]. This phenomenon is interesting for potential novel phases near the magnetic quantum critical point [19].

The $d^4$ ions, specifically Ir$^{5+}$ and Ru$^{4+}$, have been proposed as a promising platform for generating $J_{\text{eff}}$ = 0 Mott insulators [19]. From their strong spin-orbit interaction $H = \lambda \sum_i \boldsymbol{L}_i \cdot \boldsymbol{S}_i$, where λ is the spin-orbit coupling constant, the $J_{\text{eff}}$ = 0 ↔ $J_{\text{eff}}$ = 1 gap, with Kotani's theoretical calculation estimating the gap value as λ/2 [23], which is on the order of thousands of Kelvin for Ir$^{5+}$ and Ru$^{4+}$ ions. This situation makes overcoming the gap to $J_{\text{eff}}$ = 1 condensation challenging due to the extremely large λ-magnitude. Examples include the Ruddlesden-Popper type perovskite Ca$_2$RuO$_4$, identified as a $J_{\text{eff}}$ = 0 Mott insulator with magnetic ordering at ~100 K [24-26]. In Ca$_2$RuO$_4$, the observation of Higg mode is a shred of evidence of a "soft" moment based on $J_{\text{eff}}$ = 0 physics [27,28]. However, an open question is whether $J_{\text{eff}}$ = 1 condensation serves as a plausible mechanism for the softening in a magnetic moment to realize the Higgs mode remains unresolved, as the magnitude of the exchange interaction appears insufficient to overcome the $J_{\text{eff}}$ = 0 ↔ $J_{\text{eff}}$ = 1 gap in

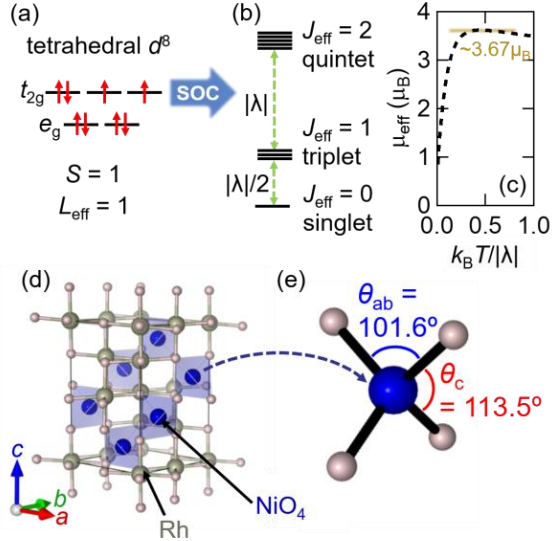

**Fig. 1** (a) Electron configuration of tetrahedrally-coordinated $Ni^{2+}$ ($d^8$) ions in $NiRh_2O_4$. (b) Energy levels scheme of $d^8$ ion under tetrahedral crystal field and spin-orbit coupling. (c) The theoretical effective magnetic moment $\mu_{eff}$ as a function of $k_B T/|\lambda|$ in $Ni^{2+}$ ion with tetragonal crystal fields, where $k_B$ is the Boltzmann constant. (d) Crystal structure of $NiRh_2O_4$ and (e) its local environment around $Ni^{2+}$ ions. The VESTA program is used for visualization [30].

ruthenate.

Very recently, it was proposed that even tetrahedrally-coordinated $Ni^{2+}$ ions with relatively weak spin-orbit coupling are a viable platform of the $J_{eff} = 0$ physics [29]. As shown in Fig. 1(a), in the tetrahedrally coordinated $3d^8$-electron configuration, the lower energy $e_g$ orbital is filled with four electrons, and residual four $d$ electrons are located in the higher-energy triply-degenerated $t_{2g}$ orbitals. As the fully-filled $e_g$ manifold can be neglected, the local physics is equivalent to the degeneracy of the $t_{2g}$ orbital in the octahedrally coordinated low-spin $d^4$-electron configuration. Therefore, as a result of SOC entangling the spin and the orbitals, a total moment $J$ is generated in the single-ion limit, with the ground state being a spin-orbital singlet with $J_{eff} = 0$ and the excited states being triplet $J_{eff} = 1$ and quintet $J_{eff} = 2$, as shown in Fig. 1(b). In this case of $Ni^{2+}$ ions, the $J_{eff} = 0 \leftrightarrow J_{eff} = 1$ gap of $|\lambda|/2 \sim 162$ cm$^{-1}$ $\sim 233$ K should compete with the superexchange interaction, and this gap is further narrowed by covalent bonding with anions and the low symmetry crystal field, which results in a gap-overcoming to $J_{eff} = 1$ condensation.

A candidate material for such a tetrahedrally coordinated $3d^8$ electron configuration is the spinel oxide $NiRh_2O_4$ [31-34]. As shown in Fig. 1(d), the $Ni^{2+}$ ions are coordinated to the tetrahedrally coordinated site of the spinel structure and the nonmagnetic $Rh^{3+}$ to the octahedrally coordinated site, resulting in a crystallization of $NiRh_2O_4$. First-principles calculations suggest that $NiRh_2O_4$ is a Mott insulator assisted by spin-orbit couplings, and the $J_{eff} = 0$ ground state is expected to be realized [32]. Furthermore, these calculations suggest the existence of a strong superexchange interaction among $Ni^{2+}$ ions through nonmagnetic $Rh^{3+}$ ions [32]. Initial reports indicated that $NiRh_2O_4$ displayed magnetic ordering at 18 K in 1963 [33]. However, subsequent research yielded contradictory findings, experimentally demonstrating that $NiRh_2O_4$ exhibits no magnetic ordering [31] and suggesting that the chemical disorder stabilizes the magnetic ordering [34]. Despite this, it is widely recognized that chemical disorder typically destabilizes magnetically ordered states, making this assertion a matter of ongoing controversy and investigation. The underlying mechanism for the observed nonmagnetic state in $NiRh_2O_4$ is the spin frustration effect arising from the diamond lattice of Ni ions. However, it is suggested that the frustration present in the material is partially mitigated by intrinsic Jahn-Teller distortion; thus, the smoking gun of this effect remains unresolved.

The magnetic order-disorder conundrum in $NiRh_2O_4$ is similar to that in $CoAl_2O_4$ with the same spinel structure [35]. The chemical formula for the spinel structure is represented by utilizing the antisite disorder parameter $\eta$, yielding $(Co_{1-\eta}Al_\eta)^{tet}[Al_{2-\eta}Co_\eta]^{oct}O_4$, in which the "tet" and "oct" superscripts refer to tetrahedral and octahedral coordination sites, respectively. The $\eta \sim 0.02$ sample shows an antiferromagnetic transition at $T_N = 6.5$ K [36], while the $\eta \sim 0.04$ sample exhibits characteristics indicative of a spin liquid-like state [37].

**TABLE I** Chemical compositions of $NiRh_2O_4$ prepared via a metathesis reaction at each of the measurement spots in Figure 2(c) and their average.

|  | #1 | #2 | #3 | #4 | #5 | Average | Ideal |
|---|---|---|---|---|---|---|---|
| NiO | 23.37 wt% | 23.99 wt% | 22.19 wt% | 21.44 wt% | 23.19 wt% | 22.8(9) wt% | 22.74 wt% |
| $Rh_2O_3$ | 78.91 wt% | 78.97 wt% | 80.24 wt% | 80.54 wt% | 79.78 wt% | 79.7(7) wt% | 77.26 wt% |
| Total | 102.28 wt% | 102.97 wt% | 102.43 wt% | 101.98 wt% | 102.97 wt% | 102.5(4) wt% | 100 wt% |
|  | apfu | apfu | apfu | apfu | apfu | apfu | apfu |
| Ni | 1.00 | 1.00 | 0.95 | 0.93 | 0.99 | 0.98(3) | 1 |
| Rh | 2.00 | 1.98 | 2.03 | 2.05 | 2.01 | 2.01(3) | 2 |
| O= | 4 | 4 | 4 | 4 | 4 | 4 | 4 |

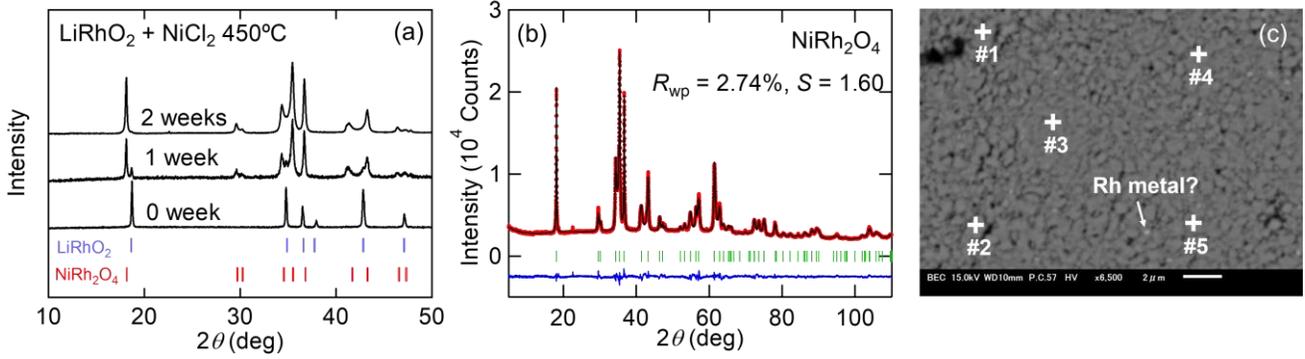

**Fig. 2** (a) X-ray powder diffraction patterns collected from reactions between LiRhO$_2$ and NiCl$_2$ at 450 °C, as a function of reaction time. Samples have been washed to remove excess NiCl$_2$. (b) Rietveld analysis results on 2 weeks sample. Observed intensities (red), calculated intensities (black), and their differences (blue) from the Rietveld refinement. Vertical bars indicate the positions of Bragg reflections. (c) A back scattered electron image of the polycrystalline sample of NiRh$_2$O$_4$ prepared via SSM. Numbers indicates EDX measurement spots.

Furthermore, the $\eta \sim 0.08$ sample exhibits behavior consistent with that of a spin glass [38]. Thus, cation-mixing is a prevalent phenomenon in spinel compounds. Similarly, the observed magnetic or nonmagnetic ground states in NiRh$_2$O$_4$ may also stem from cation-mixing, as evidenced by the detection of chemical off-stoichiometry in the crystalline structure of Ni$_{0.96}$Rh$_{1.9}$Ni$_{0.1}$O$_4$, which despite the dominance of ferromagnetic interactions, does not exhibit magnetic ordering [31]. Therefore, it is considered that undetected cation-mixing, obscured by analytical uncertainties, undermines the stability of the magnetically ordered state even in NiRh$_2$O$_4$, where it was previously assumed to be devoid of cation-mixing.

In general, at higher synthesis temperatures, the cation-mixed crystallization is more stabilized by the increase in entropy derived from cation mixing than by the loss of crystal formation energy due to cation mixing. Indeed, it has been reported that NiRh$_2$O$_4$ is synthesized at a very high temperature of 1050°C in O$_2$ flow [31]. Therefore, there is a need to develop a method to synthesize NiRh$_2$O$_4$ at lower temperatures.

In this paper, we report the observation of apparent magnetic ordering at ~45 K in NiRh$_2$O$_4$ prepared using kinetically controlled low-temperature solid-state metathesis (SSM), contrasting to the nonmagnetic behavior in the previously reported sample prepared via a high-temperature ceramic method. The observed substantial magnetic moment of ~3.7 $\mu_B$ is well accounted for by invoking the spin-orbit–entangled $J_{eff} = 0$ physics. Thus, the observed magnetic ordering is possibly associated with the condensation of $J_{eff} = 1$ exciton, driven by the interplay of the tetragonal crystal field and superexchange interactions.

## II. Experimental Methods

Following the previous report, we obtained the precursor LiRhO$_2$ by conventional solid-state reactions [39]. Stoichiometric amounts of Li$_2$CO$_3$ and Rh$_2$O$_3$ were mixed, and the mixture was calcined at 950°C for 12 h in air. Then, an SSM was performed by reacting LiRhO$_2$ with NiCl$_2$. In order to improve the reactivity, a 5:1 molar ratio of NiCl$_2$/LiRhO$_2$ was ground well together in an Ar-filled glovebox. The mixture was pelletized, sealed in a Pyrex ampoule, and then heated at 450ºC for two weeks. After the reaction, LiCl produced in the reaction and excess NiCl$_2$ were removed from samples by washing in a hot, weakly acidic aqueous solution adjusted with NH$_4$Cl and then distilled hot water repeatedly. Products were then filtered and dried at room temperature. The obtained polycrystalline samples were characterized by powder x-ray diffraction (XRD) experiments in a diffractometer with Cu-K$\alpha$ radiation , and chemical analysis was conducted using a scanning electron microscope (JEOL IT-100) equipped with an energy dispersive X-ray spectroscope (EDX with 15 kV, 0.8 nA, 1μm beam diameter). The cell parameters and crystal structure were refined by the Rietveld method using Z-Rietveld software [40].

The temperature dependence of the magnetization was measured under several magnetic fields using the magnetic property measurement system (MPMS; Quantum Design) at ISSP, the University of Tokyo. The temperature dependence of the heat capacity was measured using the conventional relaxation method in a physical property measurement system (PPMS; Quantum Design) at ISSP, the University of Tokyo. To estimate lattice contributions, we measured the specific heats of nonmagnetic spinel ZnRh$_2$O$_4$ synthesized using a conventional solid-state method.

**TABLE II** Crystallographic parameters for metathesis-prepared NiRh$_2$O$_4$ ($I4_1/amd$) determined from powder x-ray diffraction experiments. The obtained lattice parameters are $a$ = 5.9320(8) Å and $c$ = 8.743(1) Å. $B$ is the atomic displacement parameter.

| atom | Site | x | y | z | B (Å$^2$) |
|---|---|---|---|---|---|
| Ni | 4a | 0 | 1/4 | 1/8 | 1.81(3) |
| Rh | 8d | 0 | 0 | 0 | 0.73(3) |
| O | 16h | 0 | −0.0217(3) | 0.2389(2) | 1.31(5) |

## III. Results

Figure 2(a) shows powder XRD patterns of LiRhO$_2$ + NiCl$_2$ mixtures heated at 450°C over 1~2 weeks. Excess NiCl$_2$ and byproduct LiCl were removed. A slow transformation from layered rocksalt LiRhO$_2$ to spinel NiRh$_2$O$_4$ was detected. One week was not enough of a reaction; two weeks were needed for completeness. On the other hand, the XRD pattern collected from reaction mixtures heated at 400°C showed no evidence of reaction. For the sample heated at 450°C for two weeks, all the peaks can be indexed to the reflections based on the space group of $I4_1/amd$, which is the tetragonally distorted spinel structure, with the tetragonal lattice constants $a$ = 5.9320(8) Å and $c$ = 8.743(1) Å. These thus-obtained values differ from previous reports ($a$ = 5.912 Å, $c$ = 8.670 Å [31]), suggesting it is more compressed in the $c$-axis direction. This difference in the degree of crystal distortion may be due to differences in the thermodynamic or kinetic processes. Figure 2(c) presents a back scattered electron image of a post-wash powder sample of NiRh$_2$O$_4$ prepared via SSM, showcasing the fine grain size within the entire sample. The back scattered electron image also reveals tiny impurity spots, speculated to be nanosized Rh metal particles, beyond the detection limit of XRD.

To clarify the chemical composition of the prepared NiRh$_2$O$_4$ via SSM, we employed EDX analysis, the results of which are presented in TABLE I. Note that while measuring, we avoided the impurity spots mentioned above to the best of our ability, aiming to clarify a true composition. The ideal formula NiRh$_2$O$_4$ requires NiO 22.74 wt.%, Rh$_2$O$_3$ 77.26 wt.%, and total 100 wt.%. The analysis revealed minor variations in the composition, with a formula of Ni$_{0.98(3)}$Rh$_{2.01(3)}$O$_4$ based on O = 4, which aligns with the ideal NiRh$_2$O$_4$ formula within a standard deviation. In addition, the total mass percentage is near 100 wt.%, suggesting an ideal composition in the NiRh$_2$O$_4$ sample prepared via SSM without residual Li ions.

The preliminary structure designated for the Rietveld refinement was set as (Ni$_{1-\eta}$Rh$_\eta$)$_{tet}$[Rh$_{2-\eta}$Ni$_\eta$]$_{oct}$O$_4$, considering the cation mixing parameter $\eta$ at both the tetrahedral and octahedral sites as a free parameter. Intriguingly, this parameter converges to $\eta = 0$ within a standard deviation, indicating no cation mixing. The occupancy of the oxygen sites was also refined as a free parameter, but interestingly, it converged to an occupancy value of 1 within the standard deviation. Further Rietveld analysis using the initial structural model Ni$_{1-x}$Li$_x$Rh$_2$O$_4$ considering residual Li as another potential chemical disorder, also converges to the ideal chemical composition, i.e., $x = 0$ in a standard deviation. These results align with the SEM-EDX mass concentration findings. The crystallographic parameters, refined from this analysis, are detailed in TABLE II, and the corresponding visualized crystal structure is depicted in Fig. 1(d). The bond valence sum calculation for Ni and Rh ions, based on the refined structural parameters, yields +1.869 and +2.941, respectively. The nearly ideal trivalent value is estimated for Rh ions, while the slight decrease for Ni from its expected divalent state is probably due to its covalency with oxygen.

Depending on the degree of tetragonal distortion, the O-Ni-O bond angles either contract or extend from the ideal tetrahedral angle of 109.5º. The two distinct bond angles, deduced from the refined structural parameters and classified as horizontal $\theta_{ab}$ and vertical $\theta_c$ [as demonstrated in Fig. 1(e)], are observed to be 101.6º and 113.5º, respectively. These findings imply that the Ni ions are exposed to an extended tetragonal crystal field, a supposition congruent with the overarching crystal structure. Moreover, these findings align well with the previous reports [31].

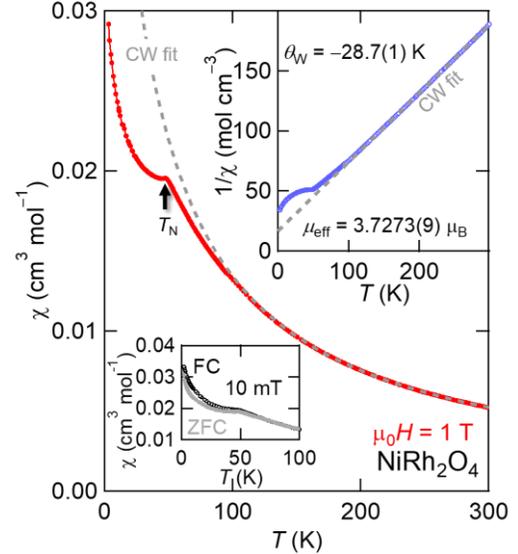

**Fig. 3** Temperature dependence of the magnetic susceptibility $\chi$ (main panel) and inverse magnetic susceptibility $1/\chi$ (right inset) for NiRh$_2$O$_4$. The gray lines on the $\chi$ and $1/\chi$ data represent fits to the Curie–Weiss (CW). The vertical arrow indicates the magnetic transition temperature $T_N$. The left inset shows the ZFC and FC $\chi$ curves of NiRh$_2$O$_4$ under low magnetic field of 10 mT.

Figure 3 shows the temperature dependence of magnetic susceptibility $\chi$ (main panel) and its inverse $1/\chi$ (right inset) for NiRh$_2$O$_4$ prepared via SSM measured at $\mu_0 H = 1$ T. A Curie-Weiss fitting of the inverse susceptibility at 200–300 K yields an effective magnetic moment $\mu_{eff} = 3.7273(9)$ $\mu_B$ and Weiss temperature $\theta_W = -28.7(1)$ K. The thus-obtained effective moment is more significant than the spin-only value of 2.828$\mu_B$ in $S = 1$, indicating the orbital contribution. Surprisingly, the observed $\mu_{eff}$ value is well matched to the fully spin-orbit-coupled value of 3.67 $\mu_B$ with $S = 1$ and $L_{eff} = 1$ calculated by Kotani as shown in Fig. 1(c) [23], indicating a realization of the spin-orbit-coupled state of Ni$^{2+}$ in NiRh$_2$O$_4$ as shown in Fig. 1(b). The negative value of $\theta_W$ indicates predominantly antiferromagnetic interactions among the Ni spins. The $\mu_{eff}$ value in Ni$_{0.96}$Rh$_{1.9}$Ni$_{0.1}$O$_4$, high-temperature NiRh$_2$O$_4$ (with potential stoichiometric deviation), and our low-temperature SSM-derived NiRh$_2$O$_4$ sample are 2.81 $\mu_B$, 3.29 $\mu_B$, and 3.72 $\mu_B$, respectively [31]. Correspondingly, the $\theta_W$ values are 24.3 K, −11.3 K, and −28.8 K, respectively [31]. This systematic variation is highly consistent when evaluated through the thermodynamic perspective of the Gibbs free energy $G = H - T\Delta S$, where the crystals containing cation-mixing are stabilized by the entropic enhancement originating

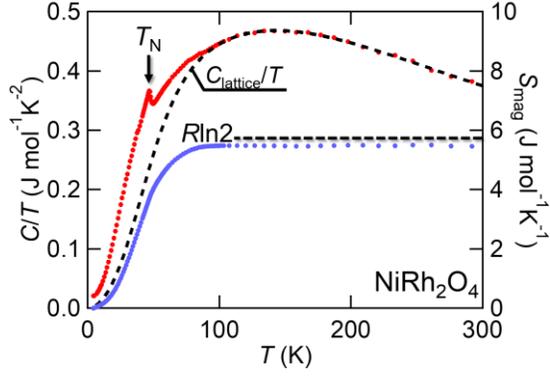

**Fig. 4** Temperature dependence of the heat capacity divided by temperature $C/T$ and estimated magnetic entropy $S_{mag}$ for $NiRh_2O_4$. The dashed line represents a lattice contribution estimated by fitting the data above 100 K ($C_{lattice}/T$). The vertical arrow indicates the magnetic transition temperature $T_N$.

from the high-temperature synthesis process.

With decreasing temperature, the $M/H$ curve gradually deviates from the Curie-Weiss fit below approximately 100 K. At lower temperatures, the susceptibility shows an apparent anomaly at $T_N \sim 45$ K, a typical behavior of antiferromagnetic ordering. Below $T_N$, $M/H$ exhibits a minor splitting of ZFC and FC curves under 10 mT, as shown in the left inset of Fig. 3, indicating the occurrence of magnetic ordering below $T_N$.

The magnetic ordering at temperatures above the absolute Weiss temperature suggests the coexistence of antiferromagnetic and ferromagnetic interactions. To confirm these expectations, we aim to roughly estimate the value of $J_2/J_1$ using a molecular field approximation. The Néel temperature $T_N$ in the $J_1$-$J_2$ diamond lattice magnets is expressed as,

$$T_N = \frac{2S(S+1)}{3k_B}(-z_1 J_1 + z_2 J_2),$$

,where $z_1$ and $z_2$ are the number of $J_1$ and $J_2$ bonds, respectively. For the diamond lattice, $z_1 = 4$ and $z_2 = 12$. This relationship indicates that the antiparallel arrangement of spins between the $J_1$ bonds in the Néel order necessarily leads to a ferromagnetic arrangement between the $J_2$ bonds. Thus, if $J_2$ is ferromagnetic ($J_2 > 0$), the Néel order is more stabilized.

On the other hand, the Weiss temperature $\theta_W$ is expressed as,

$$\theta_W = \frac{2S(S+1)}{3k_B}(z_1 J_1 + z_2 J_2).$$

The relationship between $\theta_W$ and $T_N$ yields $J_2/J_1 \sim -0.074$, indicating that the $J_2$ bond is weakly ferromagnetically coupled. Consequently, the spin model related to $NiRh_2O_4$ reveals that the $J_2$ ferromagnetism bolsters the Néel order. This inference suggests that $NiRh_2O_4$ is not spin-frustrated, thus implying a robust Néel order.

Figure 4 shows the heat capacity divided by the temperature $C/T$. An apparent λ-shaped peak is observed at 45 K. Therefore, in $NiRh_2O_4$ prepared via SSM, a second-order magnetic phase transition with bulk nature occurs at $T_N$. In order to appropriately quantify the entropy release associated with the long-range ordering in $NiRh_2O_4$, it is crucial to differentiate the magnetic heat capacity $C_{mag}$ from the lattice heat capacity $C_{lattice}$. This process involves subtracting $C_{lattice}$ from the total heat capacity. Identifying a suitable nonmagnetic reference compound for $NiRh_2O_4$ presents a significant challenge, given that it should ideally share a similar tetragonally distorted spinel structure. This structural requirement complicates matters as such a compound is not readily available. However, $NiRh_2O_4$ is characterized by a relatively small absolute Curie-Weiss temperature $|\theta_{CW}|$ without spin-frustration, and thus, $C_{mag}$ is only discernible at rather low temperatures. This characteristic allows for a broad high-temperature range to estimate $C_{lattice}$.

$C_{lattice}$ is known to consist of specific phonon contributions: three acoustic phonon branches as described by a Debye model ($C_D$) and ($3n - 3$) optical phonon branches as described by an Einstein model ($C_E$). Here, $n$ denotes the number of atoms per formula unit. For spinel compounds like $NiRh_2O_4$, $n = 7$. Given this information, we can apply a simplification procedure established in the literature [41]. $C_{lattice}$ is then simplified as,

$$\begin{aligned}C_{lattice} &= C_D + C_E \\ &= 3\left(3R\left(\frac{T}{\theta_D}\right)^3 \int_0^{\frac{\theta_D}{T}} \frac{x^4 \exp(x)}{[\exp(x)-1]^2} dx\right) \\ &+ 18R \frac{(\theta_E/T)^2 \exp(\frac{\theta_E}{T})}{\exp(\frac{\theta_E}{T}-1)},\end{aligned}$$

where $R$ is the gas constant. This model provides a satisfactory fit to the $C_p$ data for $NiRh_2O_4$ above 100 K, as depicted in Fig. 4. From this fitting, we obtain a Debye temperature ($\theta_D$) of 365(2) K and an Einstein temperature ($\theta_E$) of 707(9) K. It is important to highlight that while the Debye-Einstein model is entirely empirical, the parameters derived from the fit may not directly convey physical implications. However, it is worth noting that the obtained values are neither unrealistic nor atypical for inorganic compounds.

The magnetic contribution $C_{mag}/T$ was obtained by subtracting this lattice contribution determined by $C/T$ of the estimated $C_{lattice}/T$. Then, the magnetic entropy $S_{mag}$ was calculated by integrating $C_{mag}/T$ with respect to $T$. As shown in Fig. 4, The obtained $S_{mag}$ approaches $R\ln2 = 5.76$ J mol$^{-1}$K$^{-1}$ expected for the total entropy of spin 1/2, which is smaller than the $S_{mag}$-value of $R\ln3$ expected for the $S = 1$ spin in a $Ni^{2+}$ ion without orbital contribution. The origin of this will be discussed later based on spin-orbit coupling.

**IV. Discussion**

As described above, a magnetically ordered state was observed in $NiRh_2O_4$ prepared via SSM. Furthermore, the observed significant effective magnetic moment possibly indicates a spin-orbit entangled state rather than a trivial spin-only $S = 1$ state in tetrahedrally-coordinated Ni$^{2+}$ ions.

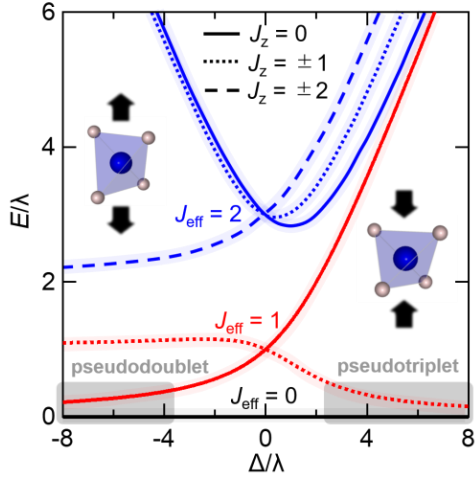

**Fig. 5** Splitting of tetrahedrally coordinated $d^8$ levels in a tetragonal crystal field; elongation and compression of the NiO$_4$ tetrahedron selects a single $J = 0$ or $J = 1$ pair state among the triplet excitations. The black, red, and blue lines correspond to energy levels derived from the $J = 0, 1$, and 2 states of the cubic condition, respectively (corresponding to Fig. 1(b)). The solid, dotted, and dashed lines indicate the $J_z = 0, 1$, and 2 states, respectively. In conjunction with the $J = 0$ ionic ground state, these magnetic states constitute a local foundation for effective low-energy models with pseudodoublet (for $\Delta/\lambda < 0$) or pseudotriplet (for $\Delta/\lambda > 0$) configurations.

We discuss a spin-orbit-entangled state in NiRh$_2$O$_4$ prepared via SSM. The enhanced effective magnetic moment confirms strong $S = 1$ and $L_{eff} = 1$ coupling. In Ni$^{2+}$ ions, the $J_{eff} = 0 \leftrightarrow J_{eff} = 1$ energy gap in the order of $|\lambda|/2 \sim 162$ cm$^{-1}$ ($\sim 233$ K) [42]. Thus, the electronic state at low temperatures is governed by the lowest energy $J_{eff} = 0$ singlet. Since the magnetic NiO$_4$ tetrahedra in the spinel structure are indirectly connected, exchange interaction is not usually large enough to overcome the $J_{eff} = 0 \leftrightarrow J_{eff} = 1$ gap.

However, as shown in Fig. 5, the low-symmetric crystal field $\Delta$ associated with the tetragonal distortion of NiO$_4$ splits the $J_{eff} = 1$ triplet into $J_z = 0$ singlet and $J_z = \pm 1$ doublet, resulting in a reduction of the magnetic gap [13,19,43,44]. Here, $\Delta/\lambda < 0$ means the NiO$_4$ tetrahedron is tetragonally elongated along the $c$-axis, and $\Delta/\lambda > 0$ is tetragonally-compressed. For the $\Delta/\lambda < 0$, like in NiRh$_2$O$_4$, the lower $J_z = 0$ level possibly condenses, resulting in a magnetic order. In this case, two low-energy states evolve $J_{eff} = 0$ and $J_z = 0$. Let us recall the $R\ln2$ magnetic entropy observed in NiRh$_2$O$_4$. The magnetic entropy value of $R\ln2$ is anticipated for the doublet ground state. In the case of NiRh$_2$O$_4$, two low-energy levels of $J_{eff} = 0$ and $J_z = 0$ are available for the magnetic ground state. It is conceivable that these levels may be compensated via superexchange interactions, resulting in a hybridized doublet state [13]. Therefore, the long range magnetic order at $T_N \sim 45$ K of NiRh$_2$O$_4$, characterized by a magnetic entropy of $R\ln2$, possibly suggests a spin-orbit entanglement with $J_z = 0$ condensation, which is facilitated by a tetragonally-elongated crystal field. This interpretation corresponds to the left side of Fig. 5, contrasting Ca$_2$RuO$_4$ with a $J_{eff} = 1$ condensation assisted by a compressed crystal field [25,26], corresponding to the right side of Fig 5. The spin-orbit–entangled $J_{eff}$ levels in NiRh$_2$O$_4$ will be elucidated by future resonant inelastic x-ray scattering (RIXS) experiments. Moreover, further theoretical investigations will clarify and resolve this intriguing aspect of the magnetism of this system.

The combination of these observations and speculations suggest that spin-orbit entangled excitonic magnetism could potentially be a plausible explanation. However, it's important to note that we are unable to completely discount the validity of alternative models. Therefore, it is necessary to accumulate more experimental evidence of the spin-orbit entangled excitonic magnetism in NiRh$_2$O$_4$. In order to construct a reasonable model and reveal the nature of magnetism in NiRh$_2$O$_4$, resonant inelastic X-ray scattering (RIXS) experiments are planned to clarify the spin-orbit-entangled state. Moreover, the proposed model of magnetic ordering, resulting from the $J_{eff} = 1$ condensations, would be more strongly supported if the softening of the magnetic moment could be detected through neutron diffraction and the Higgs mode through neutron scattering.

## V. Summary

Optimization by kinetic control of SSM has successfully improved the quality of spinel NiRh$_2$O$_4$ samples to the point that they exhibit magnetic ordering. Furthermore, the kinetically-prepared NiRh$_2$O$_4$ sample exhibits magnetic ordering below $T_N = 45$ K, probed by our experimental investigation combining bulk magnetic susceptibility and specific heat. This observation contrasts the apparent lack of magnetic ordering in the thermodynamically-prepared NiRh$_2$O$_4$ sample, possibly attributed to a slight chemical disorder as inevitable in high-temperature ceramic reactions. The observed enhanced effective magnetic moment, which is no longer $S = 1$ spin, is well explained by the strongly spin-orbit-entangled $J_{eff} = 0$ pseudospin comprised by $S = 1$ and $L_{eff} = 1$. The ground-state magnetism may be explained by the condensed state of the $J_{eff} = 1$ exciton assisted by the tetragonal crystal field and the superexchange interaction.

## Acknowledgment

This work was supported by Japan Society for the Promotion of Science (JSPS) KAKENHI Grant Number JP22K14002 and JP21K03441. Part of this work was carried out by the joint research in the Institute for Solid State Physics, the University of Tokyo.